\documentclass{iopart}
\usepackage{iopams}

\usepackage{bm} 
\usepackage{amssymb}
\usepackage{amsfonts}
\usepackage{mathrsfs}
\usepackage{comment}
\usepackage{graphicx}
\usepackage{multicol}
\usepackage{epsf}
\usepackage{epsfig}
\usepackage{fullpage}
\usepackage{setspace}

\newcommand{\NPinfo}[1]{}

\newcommand{\eqref}[1]{(\ref{#1})}

\newcommand{\calD}{{\mathcal{D}}}
\newcommand{\calE}{{\mathcal{E}}}

\newcommand{\bbR}{{\mathbb{R}}}

\newcommand{\bbZ}{{\mathbb{Z}}}

\newcommand{\rma}{{\mathrm{a}}}

\newcommand{\rmr}{{\mathrm{r}}}

\newcommand{\scrS}{{\mathscr{S}}}

\newcommand{\vecte}{{\mathbf{e}}}

\newcommand{\vectA}{{\mathbf{A}}}

\newtheorem{lemma}{Lemma}

\begin{document}

\title[Dynamical Casimir effect and dynamical systems]
        {Dynamical Casimir effect in a periodically changing domain: 
        A dynamical systems approach}

\author{Nikola P. Petrov}

\address{Department of Mathematics 
and Michigan Center for Theoretical Physics\\
University of Michigan, 
Ann Arbor, MI 48109-1109, USA}

\begin{abstract}
We study the problem of the behavior of a quantum 
massless scalar field in the space between 
two parallel infinite perfectly conducting plates, 
one of them stationary, the other moving periodically.  
We reformulate the physical problem 
into a problem about the asymptotic behavior 
of the iterates of a map of the circle, 
and then apply results from theory of dynamical systems 
to study the properties of the map.  
Many of the general mathematical properties 
of maps of the circle translate into properties 
of the field in the cavity.  
For example, we give a complete 
classification of the possible resonances in the system, 
and show that small enough perturbations 
do not destroy the resonances.  
We use some mathematical identities 
to give transparent physical interpretation 
of the processes of creation and amplification 
of the quantum field due to the motion 
of the boundary and to elucidate 
the similarities and the differences 
between the classical and quantum fields 
in domains with moving boundaries.  
\end{abstract}


\submitto{\JOB}




\section{Introduction}

Recently the problem of the behavior of the fields 
in a cavity with a (periodically) moving boundary 
has received significant attention.  
From a mathematical point of view, it constitutes 
an instructive example of parametrically driven system 
that exhibits interesting resonant effects.  
From point of view of physics, 
besides its fundamental importance 
as a modification of Casimir effect, 
it can be used as a model for the mechanism of 
formation of wave packet in lasers, 
processes in atomic physics, 
and even interstellar flight \cite{MaclayF04}!  
We would like to draw reader's attention 
to the reviews of Casimir effect 
by Bordag \etal\ \cite{BordagMM01}, 
the books by Milton \cite{Milton01}, 
and Mostepanenko and Trunov \cite{MostepanenkoTrunov1997}, 
and to the recent review 
by Dodonov \cite{Dodonov-review-01} 
devoted specifically to the dynamical (or nonstationary) 
Casimir effect (i.e., the quantum effect in a pulsating cavity) 
which contains more than 300 references.  

In this paper we will apply a method we have developed 
in \cite{LP99} (and generalized in \cite{PLV03}) 
to study the behavior of the classical 
electromagnetic field 
in a one-dimensional cavity with a moving wall 
by using methods of dynamical systems.  
Here we will employ this methodology to analyze 
the quantum problem.  
Our approach is based on studying 
the collective behavior 
of the characteristics of the wave equation 
by applying theory of circle maps.  
The general theorems allow us 
to predict the behavior of the system 
without solving partial differential equations.  
We show that the mechanism of the resonant amplification 
of the quantum field is the same as in the classical case 
-- it is due to Doppler effect at reflection from the moving mirror.  
In the quantum case, however, the motion of the mirror 
``creates'' new field which is then amplified 
by the Doppler effect.  
Using some mathematical identities, 
we give simple physical interpretation 
of the different contributions to the energy density.  

In the rest of the Introduction, we review very briefly 
some of the literature related to the mathematical 
and physical aspects of our approach, 
referring the reader to the review~\cite{Dodonov-review-01}.  

The mathematical theory of the solutions 
of the wave equation in presence of (periodically) 
moving boundaries in one or more spatial dimensions 
has been developed by Cooper 
\cite{Cooper80,Cooper93b,Cooper93a}, 
Cooper and Koch \cite{CK95}, 
Yamaguchi and collaborators 
\cite{Yamaguchi97,Yamaguchi98,Yam-98oper,Yamaguchi02}, 
Dittrich \etal\ \cite{DDG98}.  
In the physics literature, the classical version 
of the problem was studied 
by Dittrich \etal\ \cite{DDS94}, 
Cole and Schieve \cite{ColeSchieve1995}, 
M\`eplan and Gignoux \cite{MG96}, 
who used the geometric method 
of solving the wave equation 
(i.e., the method of characteristics), 
and recognized that under in some cases 
the field develops wave packets 
that become narrower with time.  

The foundations of the quantum theory of the problem 
of a 1-dimensional resonator with a moving wall were laid 
by Moore \cite{Moore1970}, and developed by 
Fulling and Davies \cite{FullingDavies1976,DaviesF77}.  
Dodonov \etal\ \cite{DKM90,DKN93a,DK96} 
considered the case of resonant motion of the mirror
within Moore's formalism; 
they and Jaekel and Reynaud \cite{JaekelR92}, 
M\'eplan and Gignoux \cite{MG96}, and others 
predicted that the force between the mirrors 
can be enhanced significantly in the resonant case.  

In \cite{Law1994_PRL}, 
Law proposed an exact analytic solution 
for a particular choice of the motion of the mirror, 
for which the field in the cavity 
develops two wave packets which become narrower in time 
and whose energy grows.  
Law's solution was generalized by Ying Wu \etal\ \cite{YCCL99}, 
who constructed motions of the mirror 
for which the field develops several wave packets.  

Dodonov \cite{Dod98} 
noticed that small enough ``detuning'' 
from the exact resonance conditions 
does not change the qualitative features 
of the behavior of the field in the cavity.  
Within our approach, one can find explicitly 
for what detuning the behavior of the field 
will change dramatically.  
The packet formation in a resonantly pulsating resonator 
was studied in detail by 
Dodonov and Andreata \cite{DA99,AD00}.  


Ideas from dynamical systems 
(i.e., study of the iterates of a certain map) 
have been used by 
Cooper \cite{Cooper93b,Cooper93a} 
(maps of the interval), 
M\'eplan and Gignoux \cite{MG96} 
(area-preserving maps of the cylinder), 
Dittrich \etal\ \cite{DDS94,DDG98}, 
Yamaguchi \cite{Yamaguchi97,Yamaguchi98,Yam-98oper} 
(circle maps).


\section{Method of characteristics and dynamical systems} 
\label{sec:char}

In this section, we explain the physical setup, 
pose the mathematical problem, 
and explain how to analyze it.  
We recommend that the reader consult our paper \cite{LP99} 
for details.  

We will refer often to the electromagnetic (EM) field 
in the cavity (meaning the classical, not quantum, field);  
as it turns out, some aspects of the problem 
are similar in the classical and in the quantum problem case.


\subsection{Description of the physical system}  
\label{sec:phys-system}

Consider the EM field in the empty space 
(no medium, electric charges, or currents) 
between two parallel perfectly reflecting mirrors, 
one stationary at $x=0$, 
the other moving according to $x=a(t)$.  
The function $a$ must satisfy the physically natural 
conditions $a(t) > 0$ for each $t$ 
(the resonator never collapses to zero length), 
$|a'(t)| < 1$ 
(the speed of the moving mirror never exceeds the speed of light).  
To avoid technicalities, we assume that $a(t)$ 
is a smooth ($C^\infty$) function.  
We will focus on the case in which 
the motion of the mirror is 1-periodic (i.e., periodic of period 1):  
\begin{equation}  \label{eq:a-per}
a(t+1) = a(t)  \quad \mbox{ for each } t \ .
\end{equation}
Examples of such functions are 
\[
a(t) = \frac{\alpha}{2} 
        + \frac{\beta}{2\pi} \sin 2\pi t 
\ , \quad \frac{\alpha}{2} > \frac{|\beta|}{2\pi} 
\ , \ \ |\beta| < 1 \ ; 
\]
\begin{equation}  \label{eq:a-complicated}
a(t) = \frac{\alpha}{2} + \frac{\beta}{2\pi} 
        \sin \left( 2\pi t + \gamma (\sin 4\pi t)^2 \right) 
\ , \quad \frac{\alpha}{2} > \frac{|\beta|}{2\pi} 
\ , \ \ |\beta| (1+ 2|\gamma|) < 1 \ . 
\end{equation}
In our numerical simulations we will use the function 
$a(t)$ from \eqref{eq:a-complicated}.


\subsection{Boundary-value problem}  \label{subsec:BVP}

Since there are no charges and currents in the cavity, 
we impose Coulomb gauge 
$A_0=0$, $\nabla\cdot\vectA = 0$ 
on the EM 4-potential $A_\mu = (A_0,\vectA)$, 
and obtain that $\vectA$ satisfies the wave equation.  
We consider linearly polarized plane waves 
propagating in $x$ direction, so we can assume 
without loss of generality 
that the vector potential has the form 
\[
\vectA(t,x) = A(t,x) \, \vecte_y \ . 
\]
The function $A(t,x)$ satisfies the $(1+1)$-dimensional 
wave equation 
\begin{equation}  \label{eq:wave}
A_{tt}(t,x) - A_{xx}(t,x) = 0 
\end{equation} 
in the spatio-temporal domain 
$\Lambda := \{(t,x)\in \bbR^2 | 0<x<a(t), \ t>0\}$.  
The boundary conditions (BCs) come from the fact 
that in the coordinate frame instantaneously 
co-moving with the mirror, 
the tangential to the mirror component of the electric field 
must vanish at the mirror, which yields 
the ``perfect reflection'' BCs
\begin{equation}  \label{eq:BC}
A_t(t,0) = 0 \ ,
\quad
A_t(t,a(t)) + a'(t) \, A_x(t,a(t)) = 0 
\end{equation}
for each $t\geq 0$.  
Geometrically, the BCs \eqref{eq:BC} 
mean that the derivative of $A(t,x)$ along the world line 
of the mirror 
(i.e., the line $\{(t,a(t))\ |\ t\in\bbR\}$ in the 
space-time diagram) must be~$0$.  
Note that the Dirichlet BCs
\begin{equation}  \label{eq:Dir-BC}
A(t,0) = c_1 = \mathrm{const} \ , \qquad 
A(t,a(t)) = c_2 = \mathrm{const}
\end{equation}
are equivalent to \eqref{eq:BC}.  
Parenthetically, we would like to note that 
Neumann BCs are not Lorenz covariant, 
so they are not physically natural 
for the case of EM fields.  
If, nevertheless, one imposes Neumann BCs, 
the predictions of the theory are dramatically different 
from those of the ``perfect reflection'' BCs \eqref{eq:BC} 
(see \cite[Section 5.6]{PLV03} 
or Dittrich \etal\ \cite[Section 4]{DDG98}).  

In the quantum case (Moore \cite{Moore1970}), 
one obtains the boundary value problem consisting of the equation 
\eqref{eq:wave}, the Dirichlet BCs \eqref{eq:Dir-BC} 
with $c_1=c_2=0$, and initial conditions.


\subsection{Method of characteristics}  

In absence of spatial boundaries 
(i.e., if $\Lambda=\{(t,x)\in\bbR^2 | t>0\}$), 
the solution of the wave equation \eqref{eq:wave} 
with ``perfect reflection'' \eqref{eq:BC} 
or Dirichlet \eqref{eq:Dir-BC} BCs, 
and initial conditions 
$A(0,x) = \psi_1(x)$, $A_t(0,x) = \psi_2(x)$, $x\in\bbR$, 
is a superposition of waves propagating to the left 
and to the right: 
$A(t,x) = \Psi^+(x_0^+) + \Psi^-(x_0^-)$, 
where $x_0^\pm = x\pm t$, and 
$\Psi^\pm(s) = \frac12 \left[\psi_1(s)\pm 
        \int^s_c \psi_2(s')\, \rmd x\right]$ 
($c$ is an arbitrary constant, 
the same for $\Psi^+$ and $\Psi^-$).  
Geometrically, in the space-time diagram, 
the waves propagate along the {\em characteristics}, 
$x\pm t=\mathrm{const}$.  

\begin{figure}
  \centerline{\psfig{file=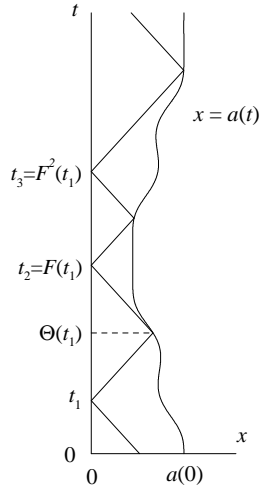,%
                     width=0.4\textwidth,angle=-90}%
        }
 \caption{\label{fig:charact}
        Characteristics of the wave equation.  
   }
\end{figure}

In the presence of spatial boundaries (stationary or moving), 
the characteristics are no more straight lines, 
but are piecewise linear, 
each part of them being a straight line at $45^\circ$ 
with respect to the $t$ axis 
(see Figure~\ref{fig:charact}).  
In order for the BCs \eqref{eq:BC} to be satisfied, 
the field changes sign at each reflection, so that 
\[
A(t,x) = (-1)^{N_+} \Psi^+(x_0^+) + (-1)^{N_-} \Psi^-(x_0^-) \ ,
\]
where $N_\pm$ is the number of reflections 
of the corresponding broken characteristic by a mirror 
between the initial moment $t=0$ and the present time $t$.  
For a proof that this algorithm works, see \cite[Section II.B]{LP99}.  


\subsection{The importance of the reflections; Doppler effect}

Loosely speaking, the ``density'' of the characteristics 
in the space-time diagram is proportional 
to the energy density of the EM field.  
Consider two characteristics corresponding to the ends 
of a narrow wave packet.  
During free propagation (no reflection), 
the width of the wave packet and the vector potential $\vectA$ 
do not change.  
At reflection from the moving mirror, however, 
not only does $A$ change sign, 
but also the width of the wave packet changes.  
If the width of the wave packet before the reflection was $\Delta$, 
and the reflection occurs at time $t$ 
(assume that the wave packet is so narrow that the reflection 
happens almost instantaneously), 
simple trigonometry shows that after the reflection 
the width $\Delta'$ of the wave packet is 
$\Delta'=\Delta/D(t)$, where 
\begin{equation}  \label{eq:D}
D(t) = \frac{1-a'(t)}{1+a'(t)} 
\end{equation}
is the {\em Doppler factor} at reflection at time~$t$.  
The term ``Doppler factor'' comes from the fact that 
the initial (classical) EM energy of the wave packet, 
$\frac{1}{8\pi}\int_\Delta 
\left[ A_t(t,x)^2 + A_x(t,x)^2 \right] \rmd x$, 
increases by a factor of $D(t)$ 
at reflection from the moving mirror 
(see \cite[Section 2.4]{PLV03} for a simple proof).  
Clearly, if the mirror is moving inwards (outwards) 
at the time of reflection, the energy of the wave packet 
will increase (decrease).  
If it happens that every time 
a certain group of nearby characteristics 
(representing a wave packet) 
is reflected from the moving mirror 
while the mirror is moving inwards, 
then they are going to get closer together, 
which will lead to 
squeezing of the wave packet 
(see Dodonov \etal\ \cite{DKM90} 
and Jaekel and Reynaud \cite{JaekelR92}) 
and to exponential growth 
of the energy of the field, as we will see below.  

Since the motion of the boundary is 1-periodic \eqref{eq:a-per}, 
the position $a(t)$ and the velocity $a'(t)$ 
of the mirror, as well as the Doppler factor 
$D(t)$ \eqref{eq:D}, 
do not depend on the integer part of $t$, 
but only on its fractional part, 
\begin{equation}  \label{eq:phase}
\{t\} := t - \lfloor t \rfloor \ , 
\end{equation} 
which we will refer to as the {\em phase} 
of the motion of the mirror.  
To make the phase of $t$ change continuously as $t$ increases, 
we will think of $\{t\}$ as belonging 
to a circle of length 1, i.e., 
to the interval $[0,1]$ with its ends identified.  
The long-time behavior of the field 
depends on the asymptotic behavior of the characteristics, 
which in turn can be analyzed by 
invoking the mathematical theory of circle maps, 
as explained below.


\subsection{From characteristics to circle maps}  

Since characteristics belong to a very simple class of plane curves 
-- namely, piecewise linear at a $45^\circ$ angle 
with the $t$ axis, -- 
to reconstruct a particular characteristic, 
it is enough to know only one moment at which 
this characteristic is reflected from, 
say, the stationary mirror.  

To study the collective behavior of the characteristics, 
we introduce the {\em time advance map} 
$F:\bbR\to\bbR$ 
such that if certain characteristic 
is reflected from the stationary mirror at time $t_1$, 
the next reflection from the same mirror occurs at time 
$t_2=F(t_1)$ (see Figure~\ref{fig:charact}).  
To derive an expression for $F$ in terms of the function 
$a$ giving the motion of the mirror, 
we notice that the time $\Theta(t_1)$ between $t_1$ and $t_2$ 
at which the characteristic is reflected from the moving mirror 
satisfies 
$\Theta(t_1) - a(\Theta(t_1)) = t_1$, 
which can be written as 
$(\mathrm{Id} - a) (\Theta(t_1)) = t_1$, 
therefore
\begin{equation}  \label{eq:Theta}
\Theta = (\mathrm{Id}-a)^{-1} \ . 
\end{equation}  
On the other hand, 
$F(t_1)=\Theta(t_1) + a(\Theta(t_1))=(\mathrm{Id}+a)(\Theta(t_1))$, 
thus 
\begin{equation}  \label{eq:F}
F = (\mathrm{Id}+a)\circ(\mathrm{Id}-a)^{-1} \ .
\end{equation} 
The conditions $a(t)>0$ and $|a'(t)|<1$ 
guarantee the invertibility of $(\mathrm{Id}\pm a)$ 
(hence the existence of $\Theta$ and~$F$) 
as well as the fact that $F$ is strictly increasing 
and, therefore, invertible.  
We leave to the reader to check that 
$a$ can be expressed in terms of $F$ as 
\[
a = \textstyle{\frac12} (F-\mathrm{Id}) \circ 
        \left[ \textstyle{\frac12}(F+\mathrm{Id}) \right]^{-1} \ .
\]

The 1-periodicity \eqref{eq:a-per} of $a$ 
guarantees that $F$ satisfies the property 
\begin{equation}  \label{eq:F-per}
F(t+1) = F(t) + 1 \ .
\end{equation}
Since only the phase $\{t\}$ \eqref{eq:phase} 
is physically important, 
instead of considering the function $F:\bbR\to\bbR$, 
we define the function 
\begin{equation}  \label{eq:f-def}
f:S^1\to S^1 : \{t\} \mapsto \{F(\{t\})\} 
\end{equation}
that maps the phase $\{t\}$ at some reflection 
from the stationary mirror to the phase $\{F(\{t\})\}$ 
at the next reflection from the same mirror.  
Here $S^1$ stands for the ``circle'', 
i.e., the interval $[0,1]$ with its ends identified 
(in mathematical notations, 
this can be written as $S^1 = \bbR/\bbZ$, 
where $\bbZ$ stands for the integers).  
The function $f$ is well-defined due to~\eqref{eq:F-per}.  

If the first reflection of a particular characteristic 
from the stationary mirror occurs at time $t_1$, 
the times of the subsequent reflections are 
$F(t_1)$, $F^2(t_1)$, $F^3(t_1)$, $\ldots$, where 
\[
F^n := \underbrace{F\circ F \circ \cdots \circ F}_{n\ \mathrm{times}} 
\]
is the $n$th {\em iterate} of the function $F$.  
Since the asymptotic behavior of the characteristics 
is completely determined by the asymptotic behavior 
of the phases at reflection, 
the long-time behavior of the system can be studied 
by analyzing the high iterates of~$f$.  
The branch of mathematics that studies 
the behavior of highly iterated functions 
is called {\em theory of dynamical systems}.  
Traditionally, the functions that are going to be iterated 
-- like $F$ and $f$ -- are called {\em maps}.  
In particular, the map $f$ \eqref{eq:f-def} 
is an example of a {\em circle map} (CM), 
i.e., a map from the circle $S^1$ to itself.  
Theory of CMs is a prominent part 
of theory of dynamical systems; 
it was initiated by Poincar\'e in 1880s, 
and nowadays is a highly developed field 
of mathematics with many physical applications.  

\begin{figure}
  \centerline{\psfig{file=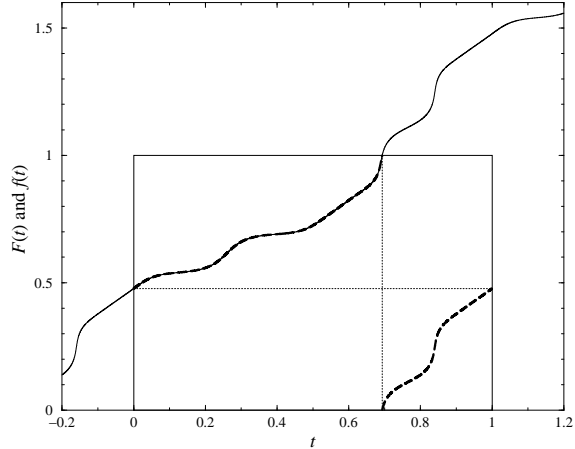,%
                     width=0.37\textwidth,angle=-90}%
        }
 \caption{\label{fig:Ff}
        Graph of the functions $F$ (thin solid line) 
        and $f$ (thick dashed line) corresponding 
        to the function $a$ \eqref{eq:a-complicated}
        for $\alpha=0.35$, $\beta=0.4$, $\gamma=0.7$.  
   }
\end{figure}

The relationship between 
the time advance map $F:\bbR\to\bbR$ 
and the CM $f:S^1\to S^1$ 
is shown pictorially in Figure~\ref{fig:Ff}.  
Note that although $f$ looks discontinuous in the figure, 
it is continuous as a function on the circle $S^1$ 
because of the identifications of $0$ and $1$ 
shown in the figure with dotted lines.  
The map $F$ is called a {\em lift} of $f$, 
while $f$ is sometimes called the {\em projection} of~$F$.  
Clearly, $F$ determines $f$ uniquely;  
on the other hand, each CM $f$ has infinitely many lifts 
that differ by an additive integer constant 
(in our case, however, the lift $F$ is defined uniquely 
by \eqref{eq:F}).  


\section{Circle maps and wave packet formation}
\label{sec:circle-maps}

In this section we collect some facts about the {\em dynamics}, 
i.e., the behavior of the high iterates $f^n$, of CMs.  
For more information the reader can consult 
the introductory expositions in Hasselblatt and Katok 
\cite[Chapter 4]{HasselblattK03} 
or Devaney \cite[Section 1.14]{Devaney89}, 
or the more sophisticated treatments in 
Katok and Hasselblatt \cite[Chapters 11 and 12]{KH95}, 
de Melo and van Strien \cite[Chapter I]{dMvS93}.  
Section III of our paper \cite{LP99} contains a selection 
of mathematical facts adapted 
to the problem of the resonator.  
In Section~\ref{subsec:maps-physics} 
we will give interpretation of the mathematical results 
in terms of the asymptotic behavior 
of the field in the resonator.  


\subsection{Circle maps -- basic definitions}  
\label{subsec:circle-defs}

By a circle map (CM), we will always mean a smooth ($C^\infty$) 
invertible map of the circle whose inverse is also smooth 
-- this is exactly the class of CMs 
that correspond to motions of the boundary $a$ 
satisfying the conditions from Section~\ref{sec:phys-system}.  
For the map $f$ to be invertible, 
we have to assume that the cavity is not too long, 
or, more concretely, that $F(t)-t < 1$ 
(for which it is enough to assume that $a(t)<\frac12$ 
for all $t$).  
This condition only helps to avoid clumsy sentences, 
but is not a restriction of the generality 
-- our ideas can be easily applied 
{\em mutatis mutandis} to the case of a longer cavity.

The most important characteristic of a CM 
is its {\em rotation number} defined 
as the ``average amount of rotation'': 
\begin{equation}  \label{eq:rot-num}
\tau(f) \equiv \tau(F) 
:= 
\lim_{n\to\infty} \frac{F^n(t)-t}{n} 
\end{equation}
(with the above restriction on the length of the cavity, 
$\tau(f)\in [0,1)$).  
It can be proved that $\tau(f)$ always exists 
and does not depend on the value of~$t$ in \eqref{eq:rot-num}.  

An {\em orbit} of a point $t\in S^1$ 
is the set $\{f^n(t)\}_{n=0}^\infty$ 
of all future (i.e., for $n\geq 0$) iterates of $t$.  
If for some point $t^*\in S^1$ 
there exists an integer $q$ such that $f^q(t^*)=t^*$, 
then we say that $t^*$ is a {\em periodic point} of period~$q$ 
(or a $q$-{\em periodic point}) 
and call the orbit $\{f^n(t^*)\}_{n=0}^{q-1}$ of this point 
a {\em $q$-periodic orbit}.  

The simplest example of a CM 
is the {\em rigid rotation} $r_\sigma:S^1\to S^1$ 
(where $\sigma\in[0,1)$) 
defined through its lift $R_\sigma$, 
\begin{eqnarray}  \label{eq:rigid}
R_\sigma(t) &:=& t + \sigma \ , \quad t \in \bbR \ ,
        \nonumber       \\[-3.5mm]
                        \\[-3.5mm]
r_\sigma(t) &:=& \{t + \sigma\} \ , \quad t\in S^1 \ . \nonumber
\end{eqnarray}
Clearly, $\tau(r_\sigma) = \sigma$.  
The dynamics of $r_\sigma$ is very simple:
\begin{itemize}
\item
if $\sigma$ is a rational number, 
i.e., $\sigma = p/q$ for some integers $p$ and $q$ 
(we will always assume that $p$ and $q$ 
do not have common factors), 
then after $q$ iterations any point $t\in S^1$ 
returns to its initial position, 
having traversed the circle $p$ times, i.e., 
each point $t\in S^1$ is a $q$-periodic point:  
\begin{eqnarray*}
R_{p/q}^q(t) &=& t + p  \quad \forall t \in \bbR \ ,
        \nonumber       \\[-3.5mm]
                        \\[-3.5mm]
r_{p/q}^q(t) &=& t  \quad \forall t \in S^1 \ ;
\end{eqnarray*}

\item
if $\sigma$ is not a rational number, 
then the orbit $\{r_\sigma^n(t)\}_{n=0}^\infty$ 
of any point $t\in S^1$ fills the circle densely, 
and will never return to the initial point $t$, 
thus, in this case there are no periodic orbits.  
\end{itemize}


\subsection{Phase locking, Arnol'd tongues, devil's staircase}  

Here we will describe in detail the case 
of a general CM $f$ with a rational rotation number, $\tau(f)=p/q$, 
in which case the map $f$ is said to be 
{\em phase locked} 
({\em frequency locked}, {\em mode locked}).  

If $\tau(f)=p/q$, then generically $f$ has 
an attracting $q$-periodic orbit $\{t_j^{(\rma)}\}_{j=1}^q$, 
and a repelling $q$-periodic orbit $\{t_j^{(\rmr)}\}_{j=1}^q$.  
``Attracting'' means that the orbit of each point $t\in S^1$ 
that is not one of the of the repelling periodic points 
$\{t_j^{(\rmr)}\}_{j=1}^q$ tends asymptotically 
to the attracting periodic orbit $\{t_j^{(\rma)}\}_{j=1}^q$.  
The repelling periodic orbit ``repels'' the iterates of $f$; 
it is an attracting periodic orbit for the inverse map $f^{-1}$ 
(which is also a CM).  
The attracting and repelling periodic orbits 
of the CM $f$ give rise to attracting and repelling 
characteristics of the wave equation, 
and to formation of wave packets 
(see Section~\ref{subsec:maps-physics}).  

\begin{figure}
  \centerline{\psfig{file=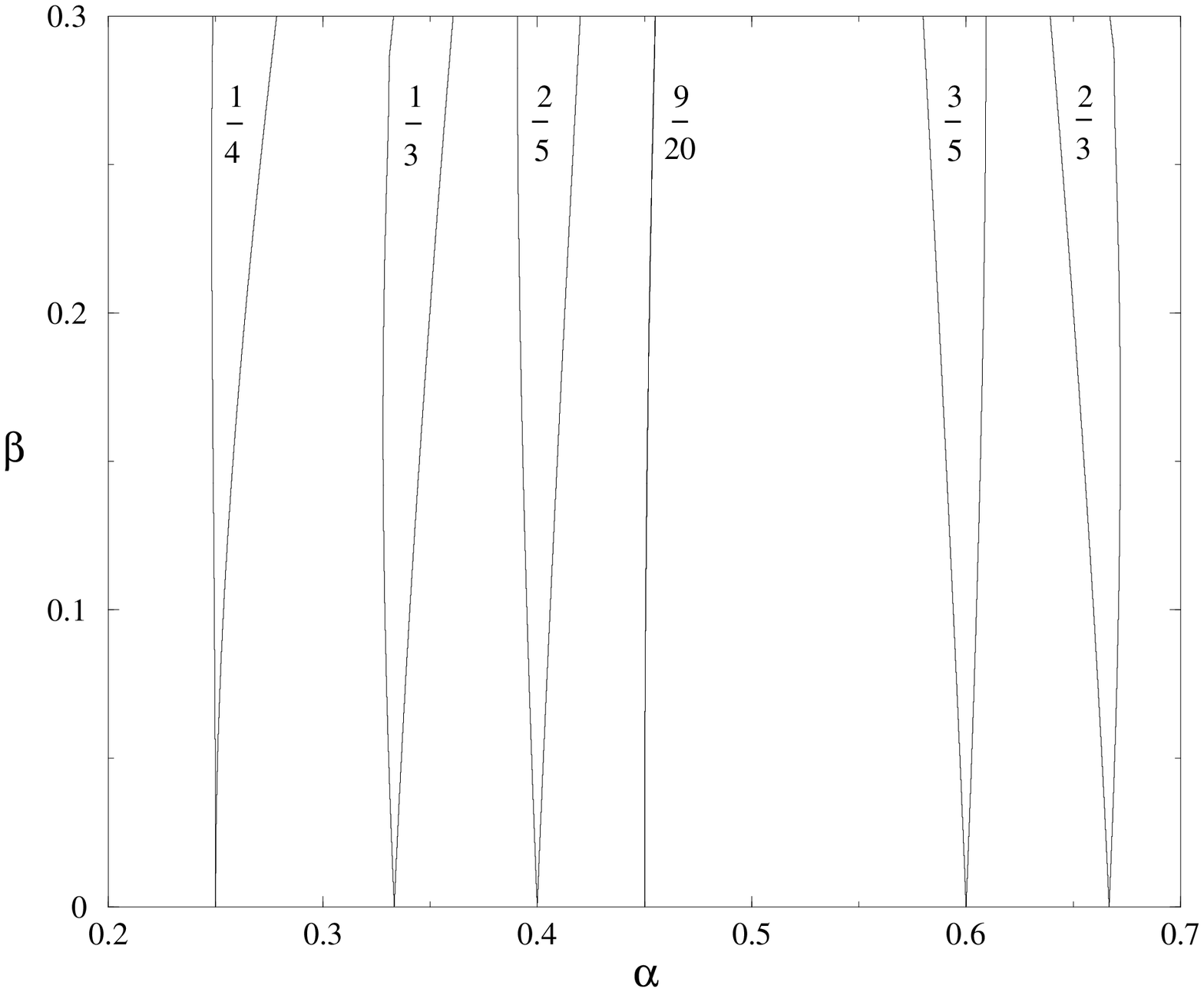,%
                     width=0.37\textwidth,angle=-90}
        \hspace{0.06\textwidth}
        \psfig{file=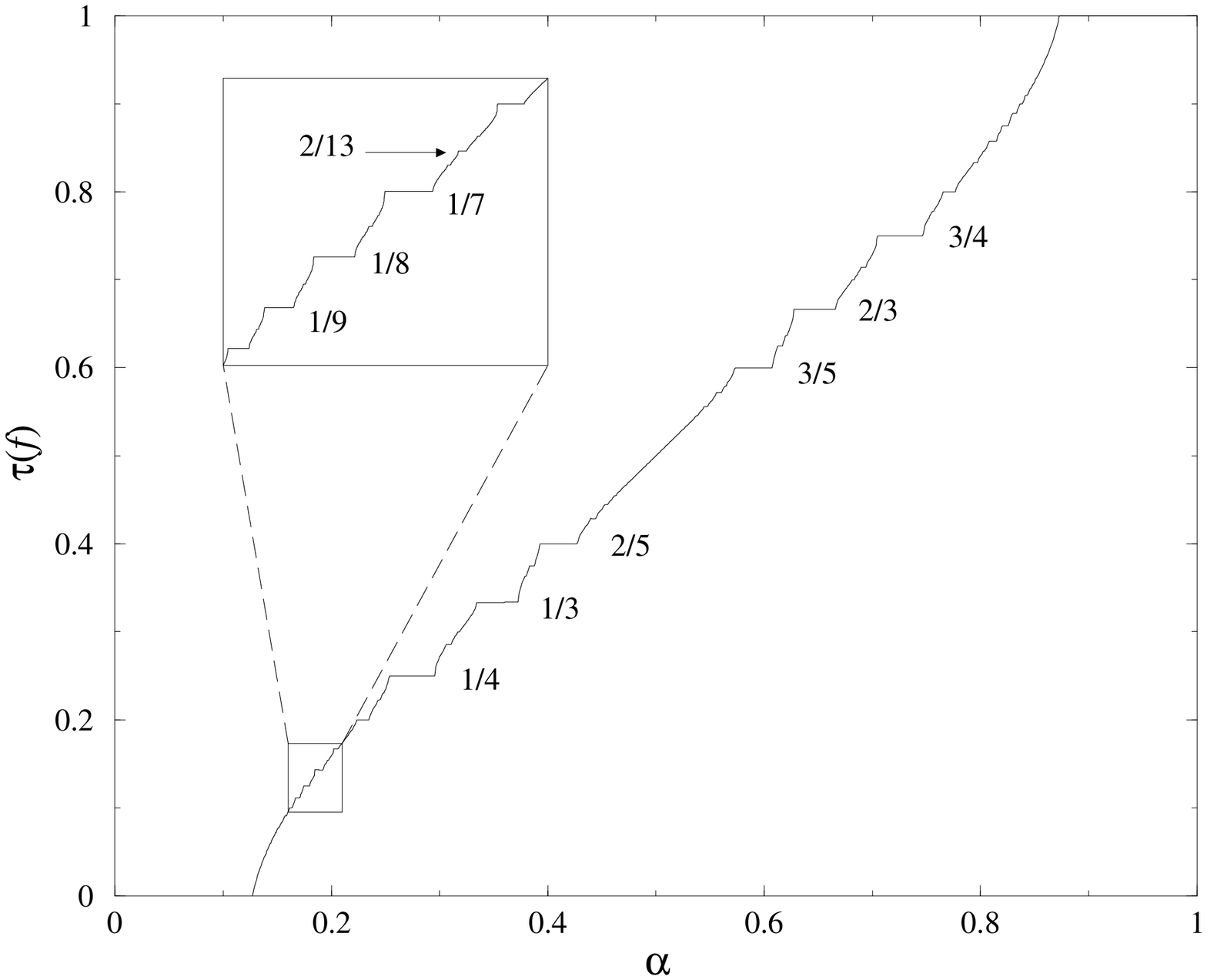,%
                     width=0.37\textwidth,angle=-90}}
 \caption{\label{fig:devil}
        (a) Arnol'd tongues with rotation numbers $1/4$, $1/3$, 
        $2/5$, $9/20$, $3/5$, and $2/3$ 
        for the CM $f$ corresponding to 
        motion of the boundary \eqref{eq:a-complicated} 
        with $\gamma=0.7$.  
        (b) Graph of the rotation number $\tau(f)$ for \eqref{eq:a-complicated} 
        with $\beta=0.4$, $\gamma=0.7$.  
   }
\end{figure}

A very important for the physics of the problem question 
is how ``generic'' the case of phase locking is.  
In Figure~\ref{fig:devil}(a) we show 
in the $(\alpha,\beta)$-plane 
the regions of values of the parameters 
$\alpha$ and $\beta$ of the motion of the mirror 
that correspond 
to phase-locking of several rotation numbers $\tau(f)$.  
These ``phase-locked'' regions in the $(\alpha,\beta)$-plane 
are called {\em Arnol'd tongues} 
in honor of Arnol'd who studied them 
in his famous paper on CMs \cite{Arnold1961}.  
The Arnol'd tongue corresponding to a $p/q$ phase locking 
emanates (i.e., starts as $\beta\to0^+$) from $\alpha=p/q$, 
and becomes thicker as $\beta$ increases.  
Tongues with large $p$ and $q$ are very thin 
(see the $9/20$ tongue in Figure~\ref{fig:devil}(a)).  

Another illustration of the abundance of phase-locking 
is the graph of the rotation number $\tau(f)$ versus $\alpha$ 
(all other parameters fixed), shown in Figure~\ref{fig:devil}(b).  
It can be proved that this function is continuous, 
and it is locally constant if $\tau(f)$ is rational, 
and strictly increasing if $\tau(f)$ is irrational, 
i.e., the graph of such a function contains 
infinitely many densely interspersed horizontal pieces, 
each corresponding to a particular type of phase locking.  
Such a graph is called a {\em devil's staircase}.  
Several of the horizontal pieces in the figure are labeled 
with $p/q$ showing the type of phase locking.  
The behavior of the widths of the Arnol'd tongues 
as $\beta\to 0^+$ is studied by Jonker \cite{Jonker90} 
and Davie \cite{Davie1996}.  

It is worth noting that, in some sense, 
phase locking is more ``generic'' than the unlocked case.  
Namely, if $\tau(f)$ is irrational, 
then there exists an arbitrarily small smooth perturbation 
of $f$ such that the perturbed map is phase locked.  
On the other hand, if $\tau(f)=p/q$, 
then the parameter values are either strictly inside 
or on the boundary of the $q/p$-Arnol'd tongue.  
If the parameters are strictly inside the tongue, 
then a small enough (but otherwise arbitrary) smooth perturbation 
will not change its rotation number, i.e., the perturbed map will 
have rotation number $p/q$.  
Practically, however, one should not forget 
that the width of the Arnol'd tongues 
decreases fast when $p$ and $q$ increase 
or when $\beta$ decreases.  

If the CM $f$ is in $p/q$ phase locking 
and the parameters are strictly inside the $p/q$ tongue, 
$f$ has an attracting $q$-periodic orbit and a repelling one.  
In this case, if $t^*$ belongs to the attracting periodic orbit, 
then $f^q(t^*)=t^*$ and $f'(t^*)<1$.  
In the following, we will say that a phase locking is ``generic'' 
if the parameters of the CM are strictly inside the Arnol'd tongue, 
in which case there exist an attracting and a repelling orbit.  
If the parameters of the map are on the boundary 
of the Arnol'd tongue (i.e., at some end on the corresponding 
horizontal piece of the graph in Figure~\ref{fig:devil}(b)), 
then there exists a $q$-periodic orbit 
which is neither attracting, nor repelling; 
in this case $f'(t^*)=1$ where $t^*$ is a $q$-periodic point.  

When $\beta$ reaches the critical value 
at which the boundary is moving at the speed of light 
at some moment in each period (i.e., if $|a'(t)|=1$ for some $t$), 
then the total length of the phase locking intervals 
(i.e., of the horizontal pieces in Figure~\ref{fig:devil}(b)) 
becomes equal to~1, 
which means physically that the probability 
of phase locking is~1.  
This mathematical problem is studied numerically 
by Jensen \etal\ \cite{Jensenetal1984} 
and Lanford \cite{Lan85}, 
and proved rigorously 
for general CMs by Graczyk and {\'S}wi{\c{a}}tek \cite{GS96}.

\subsection{Derivative of the circle map, Doppler factor, 
        formation of wave packets}  \label{subsec:maps-physics}

Now we translate the mathematical facts 
about the dynamics of the CM $f$ \eqref{eq:f-def} 
and the time advance map $F$ \eqref{eq:F} 
into asymptotic properties 
of the field in the cavity.  

If a particular characteristic 
is reflected by the stationary mirror at time $t$, 
then the times of the subsequent reflections 
from the same mirror are given by $F^n(t)$ ($n=1,2,\ldots$); 
the phases \eqref{eq:phase} of the motion of the mirror 
at these times are given by the iterates of the corresponding CM, $f^n(\{t\})$.  
If $f$ is phase locked with rotation number $\tau(f)=p/q$, 
then, generically, there exists an attracting 
periodic orbit $\{t_j^{(\rma)}\}_{j=1}^q$ 
which attracts the iterates of any point in $S^1$ 
(except the repelling periodic points) under the map~$f$.  
Physically, each attracting periodic point $t_j^{(\rma)}\in S^1$ 
of the CM $f$ 
corresponds to an infinite sequence of times of the form 
$n + t_j^{(\rma)}$, where $n$ is any integer, 
such that the characteristics that are reflected 
from the stationary mirror at these times 
attract the nearby characteristics.  
We will call times of the form $n + t_j^{(\rma)}$ 
``attracting $q$-periodic times'', 
and the corresponding characteristics 
``attracting $q$-periodic characteristics''.  
The presence of $q$ attracting periodic characteristics 
means physically that the field in the cavity 
develops (at most) $q$ wave packets, 
whose widths decrease exponentially, 
and whose energies increase exponentially with time.  

In the case of generic $p/q$ phase locking, 
the rate at which the characteristics get closer together 
is related to the first derivative of $F$ (or, equivalently, $f$), 
which in turn is related 
(according to \eqref{eq:D}, \eqref{eq:Theta}, \eqref{eq:F}) 
to the Doppler factor $D(\Theta(t))$ 
at the time $\Theta(t)$ of the first reflection 
from the moving mirror after $t$:  
\[
F'(t) = \frac{1+a'(\Theta(t))}{1-a'(\Theta(t))} 
        = \frac{1}{D(\Theta(t))}  \ .
\]
Asymptotically, the wave packets are very narrow, 
so that they are reflected 
from the moving mirror practically instantaneously, 
at times of the form $n + t_j^{(\rma)}$.  
The asymptotic ``cumulative'' Doppler factor $\calD_q$ 
over a sequence of $q$ consecutive reflections 
of the packet from the moving mirror 
(which takes total time $p$ according to the fact that 
$F^q(t_j^{(\rma)}) = t_j^{(\rma)} + p$) 
is equal to the product of Doppler factors 
at each of these $q$ reflections:  
\begin{equation}  \label{eq:D-q}
\calD_q := \prod_{j=1}^{q} D(\Theta(t_j^{(\rma)})) 
= \left[ \prod_{j=1}^{q} F'(t_j^{(\rma)}) \right]^{-1} 
= \left[ (F^q)'(t_1^{(\rma)}) \right]^{-1} 
\ . 
\end{equation}
This formula holds exactly only in the case of classical EM field, 
when the motion of the mirror only amplifies the field 
through Doppler effect at reflection.  
In Section~\ref{subsec:energy}, we will see that in the quantum case 
the energy is not only amplified, but also created by the motion 
of the mirror, which introduces corrections 
to the rate of change of the energy.  

In the case when the rotation number $\tau(f)$ is irrational, 
the field does not develop wave packets, 
and its energy changes with time, but does not have a tendency 
towards steady grow or decay.


\section{Quantum effects in a periodically pulsating resonator}
\label{sec:quantum-effects}

\subsection{Moore's functional equation}

Moore \cite{Moore1970} was the first to consider 
the problem of quantizing the electromagnetic field 
in a one-dimensional resonator with a moving wall.  
We leave out all the complications that he had to overcome 
in the development of a quantization scheme, 
and focus on one particular aspect of his treatment 
(adapting his equations to our approach).  
Let the motion of the mirror correspond to time advance map $F$ \eqref{eq:F} 
of rotation number~$\sigma$.  
Moore showed that in this case one has to look for an expansion 
of the field operator $A(t,x)$ in mode functions 
\[
A_k(t,x) = \e^{-2\pi \rmi k \frac1\sigma \Sigma(t-x)} 
        - \e^{-2\pi \rmi k \frac1\sigma \Sigma(t+x)} \ , 
\quad k = 1,2,3,\ldots \ ,
\]
where the function $\Sigma:\bbR\to\bbR$ 
satisfies Moore's functional equation 
\[
\Sigma(t+a(t)) = \Sigma(t-a(t)) + \sigma \ , 
\]
which ensures that $A_k(t,x)$ satisfy the Dirichlet 
BCs \eqref{eq:Dir-BC} with $c_1=c_2=0$.  
This equation can be rewritten in terms of the map $F$ as 
$\Sigma\circ F (t) = \Sigma(t) + \sigma$, or, equivalently, as 
\begin{equation}  \label{eq:SigmaR}
\Sigma \circ F = R_\sigma \circ \Sigma 
\end{equation}
(where $R_\sigma$ \eqref{eq:rigid} 
is the rigid rotation by~$\sigma$), 
and interpreted as the fact that the value of $\Sigma$ 
changes between two consecutive reflections 
from the stationary mirror by~$\sigma$.  
This implies, in particular, that if for some particular value 
$\bar t$ we know the values of $\Sigma(t)$ in the interval 
$[\bar t, F(\bar t))$, then we can reconstruct 
the function $\Sigma$ for all $t\in\bbR$ 
by using \eqref{eq:SigmaR}.

Moore's functional equation is easy to solve numerically.  
Let us assume that before $t=0$, the two mirrors were at rest, 
and at $t=0$ the right mirror started moving:
\begin{equation}  \label{eq:a-smooth}
a(t) = 
\left\{
        \begin{array}{ll}
        \frac{\alpha}{2} \ ,    & t \leq 0 \\
        \frac{\alpha}{2} + \frac{\beta}{2\pi} ( 1 - \e^{-t^4} ) 
        \sin \left( 2\pi t + \gamma (\sin 4\pi t)^2 \right) 
                \ ,     & t>0 \ .
        \end{array}
\right.
\end{equation}
The motion of the boundary for $t>0$ is very similar 
to the one in \eqref{eq:a-complicated}; 
we have used the factor $1 - \e^{-t^4} $ 
to smooth out the transition (i.e., to ensure 
that $\Sigma'$, $\Sigma''$, and $\Sigma'''$ are continuous, 
the reason for which will become clear 
in Section~\ref{subsec:energy}) 
the factor $( 1 - \e^{-t^4} )$ 
tends to 1 very quickly as $t$ grows, so it does not 
affect the asymptotic behavior of the system.  
\begin{figure}
  \centerline{\psfig{file=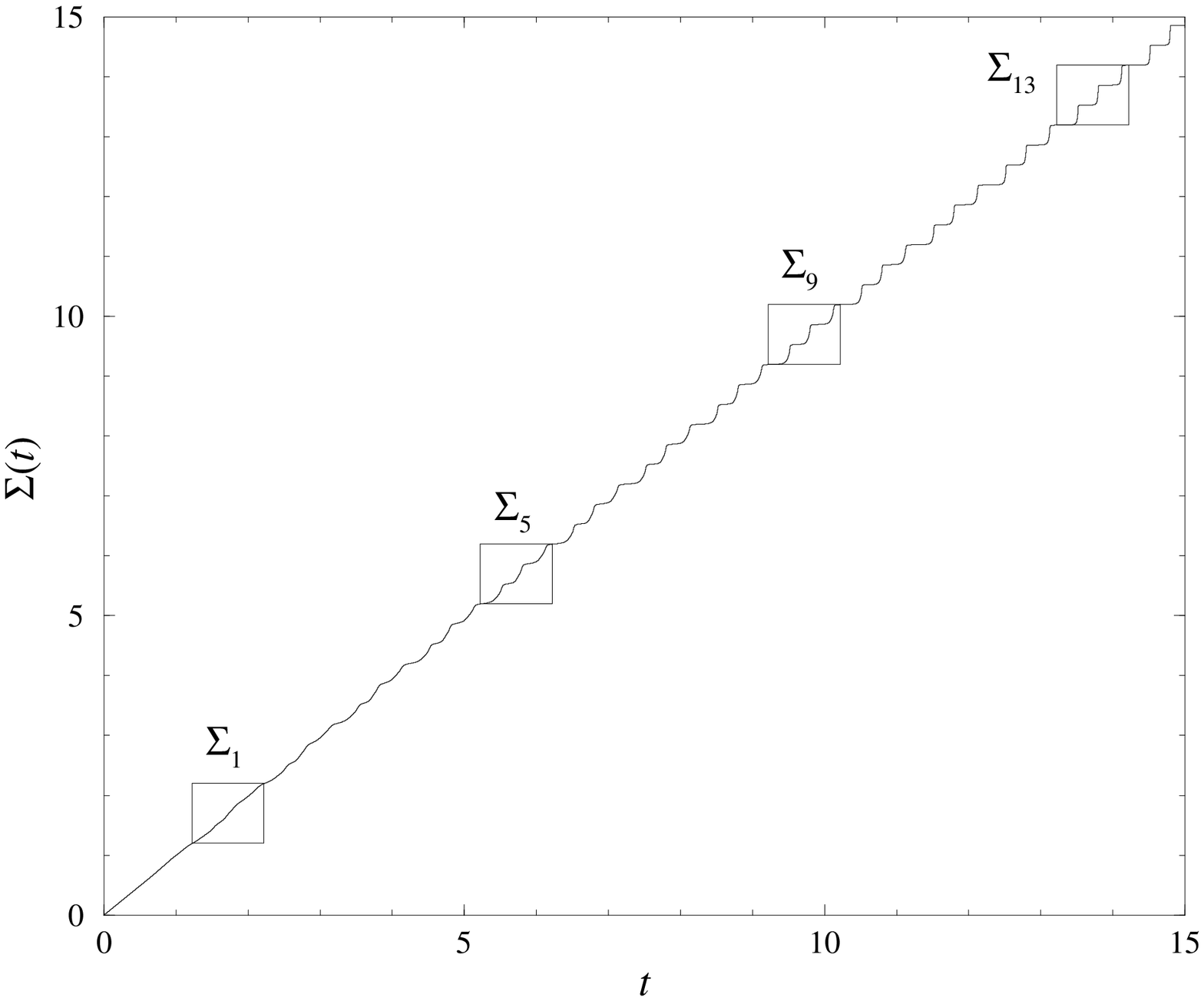,%
                     width=0.39\textwidth,angle=-90}
        \hspace{0.06\textwidth}
        \psfig{file=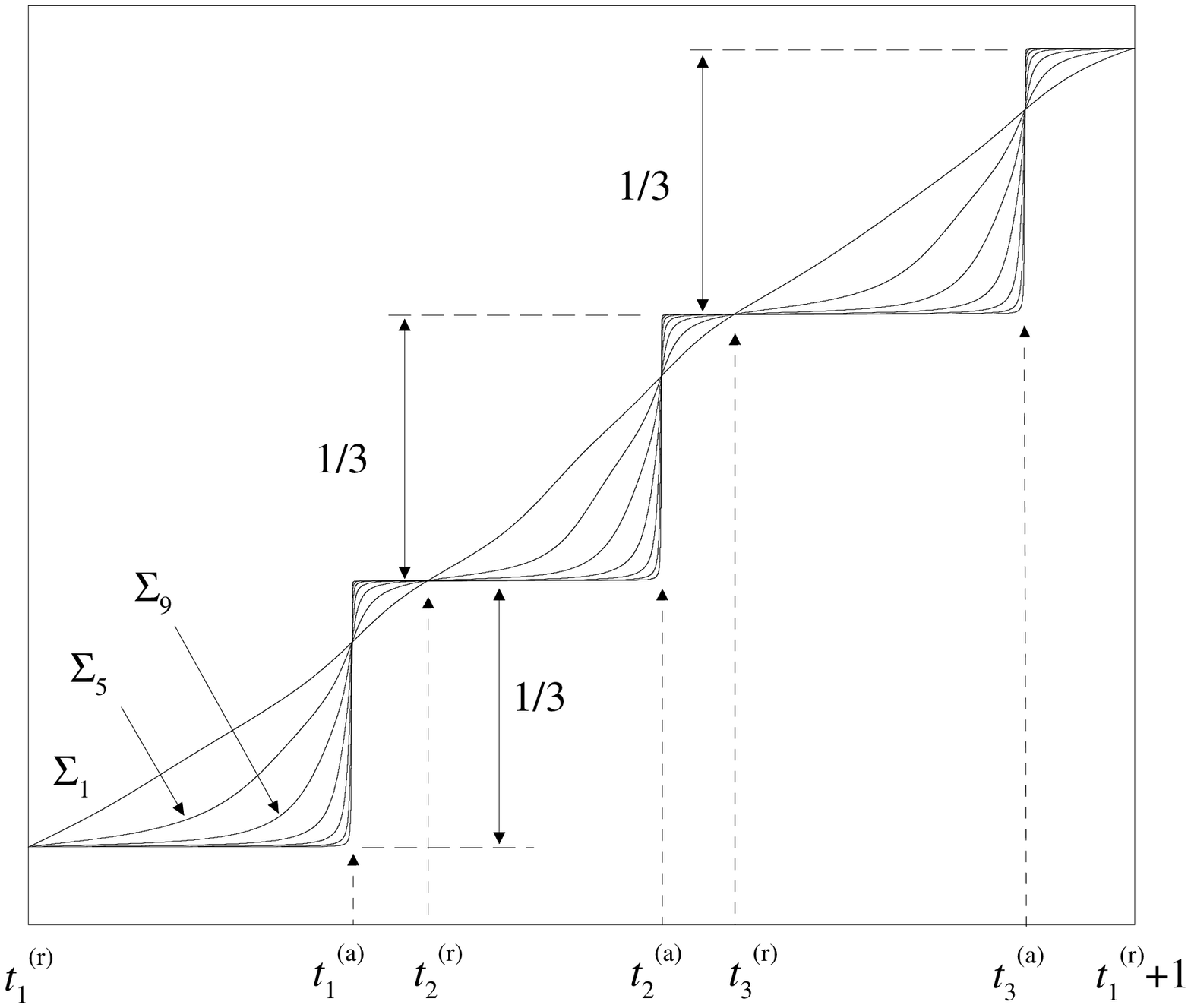,%
                     width=0.365\textwidth,angle=-90}}
 \caption{\label{fig:graph-S}
        (a) Development of the staircase-like structure 
        of the function~$\Sigma$ 
        for motion of the mirror given by 
        \eqref{eq:a-smooth} with 
        $\alpha=0.34$, $\beta=0.2$, $\gamma=0.3$, 
        corresponding to $1/3$-phase locking.  
        (b) Graphs of the functions $\Sigma_n$ \eqref{eq:Sigman}, 
        $n=1$, 5, 9, 13, 17, 21, 25, 
        for the same parameter values.  
   }
\end{figure}
%
%
%
%
If the mirrors are stationary, 
it is natural to take the function $\Sigma(t)$ to be linear, 
so for $t\in[-\alpha,0)$, 
we take $\Sigma(t) = \frac{\sigma}{\alpha}t + \mathrm{const}$ 
(the constant is immaterial since only 
the derivatives of $\Sigma$ have physical meaning), 
and then use \eqref{eq:SigmaR} to find $\Sigma(t)$ for $t\geq 0$.

Now we will apply our knowledge about 
the dynamics of the CM $f$ 
-- and, hence, about the time advance map $F$ -- 
to draw conclusions about the asymptotic behavior 
of the function $\Sigma$, which 
will allow us to make predictions 
about the long-time behavior of the energy density of the field.  
Since the case of a rational $\sigma=p/q$ is especially interesting 
because of the occurrence of resonant phenomena (phase locking), 
we focus on this case in the rest of this subsection.  
In the case of a generic $p/q$ phase locking, the CM $f$ has 
an attracting and a repelling $q$-periodic orbits.  
Let $t_1^{(\rma)},\ldots,t_q^{(\rma)}$ be a sequence of $q$ consecutive 
attractive $q$-periodic times, and let the ordering be such that 
\[
t^{(\rma)}_1 \stackrel{F}\mapsto 
t^{(\rma)}_2 \stackrel{F}\mapsto 
\cdots \stackrel{F}\mapsto 
t^{(\rma)}_q \stackrel{F}\mapsto t^{(\rma)}_1 + p 
\]
(this implies that the fractional parts of these times satisfy 
$\{t^{(\rma)}_1\} \stackrel{f}\mapsto 
\cdots \stackrel{f}\mapsto 
\{t^{(\rma)}_q\} \stackrel{f}\mapsto \{t^{(\rma)}_1\}$).  
If $n$ is a positive integer, then \eqref{eq:SigmaR} 
iterated $nq$ times reads 
\[
\Sigma \circ F^{nq} (t) = \Sigma(t) + np \ .
\]
Using that $F^{nq} (t_1^{(\rmr)}) = t_1^{(\rmr)} + np$, 
we obtain 
\[
\Sigma ( t_1^{(\rmr)} + np ) = \Sigma(t_1^{(\rmr)}) + np \ .
\]
This allows us to 
define a sequence of functions (for $n=0$, 1, $\ldots$) 
\begin{equation}  \label{eq:Sigman}
\Sigma_n : [ t_1^{(\rmr)}, t_1^{(\rmr)}+np ) \to 
                [ \Sigma(t_1^{(\rmr)}), \Sigma(t_1^{(\rmr)})+np )  
\end{equation} 
\[
\Sigma_n(t) := \Sigma(t+np) - np \ ,
\]
and study the behavior of $\Sigma(t)$ for very large $t$ 
by analyzing the behavior of $\Sigma_n(t)$ 
(where $t\in [ t_1^{(\rmr)}, t_1^{(\rmr)}+np )$) for $n\to\infty$.  
The graph of $\Sigma$ can be assembled from translates 
of the graphs of $\Sigma_n$ as shown in Figure~\ref{fig:graph-S}(a), 
for parameters corresponding to $1/3$ phase locking.  
In Figure~\ref{fig:graph-S}(b), 
we show the graphs of several~$\Sigma_n$ 
for the same parameter values.  

In the case of generic $p/q$ phase locking, the times of reflection, 
$t$, $F(t)$, $F^2(t)$, $\ldots$, of a particular characteristic 
from the stationary mirror 
accumulate at the attracting periodic times 
(of the form $n + t_j^{(\rma)}$), 
while the values of $\Sigma$ at two consecutive reflections 
differ by the constant value $\sigma$.  
This difference between the behavior of the arguments 
and the values of the function $\Sigma$ (cf.\ \eqref{eq:SigmaR}) 
explains the occurrence of exactly $q$ steep parts 
of the graphs of $\Sigma_n$ at $t = t_j^{(\rma)}$ ($j=1,\ldots,q$), 
and $q$ almost horizontal parts for large~$n$.  
Physically, the steep parts correspond to 
the times of reflection of the packets of the field; 
the ``widths'' of these packets 
are given approximately by (cf.\ \eqref{eq:D-q}) 
\[
\Delta_j^{(n)} \sim \mathrm{const} \cdot [ (F^q)'(t_1^{(\rma)})]^n 
= \frac{\mathrm{const}}{\calD_q^n} \ .
\]


\subsection{Energy of the quantum field}  \label{subsec:energy}

Fulling and Davies \cite{FullingDavies1976} 
computed the energy density of the quantum field 
in a one-dimensional cavity with one stationary 
and one moving mirror using the ``point-splitting'' method 
of DeWitt \cite{DeWitt75}.  
They found that the regularized energy density 
in the space between the mirrors 
(i.e., the energy minus an infinite constant) 
is a superposition of the energies 
of left- and right-propagating disturbances:  
\begin{equation}  \label{eq:en-dens}
\langle T_{00}(t,x) \rangle_\mathrm{reg} 
= -\frac{1}{24\pi} \left[ \Phi(t+x) + \Phi(t-x) \right] \ ,
\end{equation}
where 
\[
\Phi(\xi) 
        = \scrS_\Sigma(\xi) + \frac{2\pi^2}{\sigma^2} \left[\Sigma'(\xi)\right]^2 \ ,
\]
and $\scrS_\Sigma$ is the {\em Schwarzian derivative} 
of the function $\Sigma$, defined as 
\begin{equation}  \label{eq:Schw}
\scrS_\Sigma(z) 
        := \frac{\Sigma'''(z)}{\Sigma'(z)} 
        - \frac32\left[\frac{\Sigma''(z)}{\Sigma'(z)}\right]^2 \ .
\end{equation}
The Schwarzian derivative is a remarkable 
(highly nonlinear!)\ differential operator, 
first introduced in complex analysis.  
If $\phi$ is a complex analytic function, 
then 
vanishing of $\scrS_\phi$ is a necessary and sufficient 
condition for 
$\phi$ to be a M\"obius (i.e., fractional linear) transformation, 
$M(z) = \frac{az+b}{cz+d}$, where $ad-bc\neq0$ 
(see, e.g., Nehari \cite[Chapter V]{Nehari52}).  
The Schwarzian derivative is invariant with respect to 
a composition with a M\"obius transformation, 
$\scrS_{M\circ F} = \scrS_F$, 
which follows from the identity 
\begin{equation}  \label{eq:S-GoH}
\scrS_{G\circ H}(z) = \scrS_G(H(z)) \, \left[H'(z)\right]^2 
        + \scrS_H(z) \ .
\end{equation}
The Schwarzian derivative appears in many branches of mathematics 
-- dynamical systems 
(Singer \cite{Singer78}, 
de Melo and van Strien \cite[Chapter 1]{dMvS93}, 
Graczyk \etal\ \cite{GSS01}), 
Lorentzian geometry 
(Kostant and Sternberg \cite{KostantS88}, 
Duval and Guieu \cite{DuvalG00}, 
Duval and Ovsienko \cite{DuvalO00}, 
Singer \cite{Singer01}), 
theory of differential equations 
(Hille \cite[Chapter 10]{Hille76}), 
integrable systems 
(Burstall \etal\ \cite{BurstallPP02}), 
among many others.  
Even more interestingly, 
Schwarzian derivative is widely used 
as a tool in theory of CMs 
(Herman \cite{Her85}, 
Graczyk and {\'S}wi{\c{a}}tek \cite{GS96}).  

One can use the the property \eqref{eq:SigmaR} 
and the composition rule \eqref{eq:S-GoH} 
to predict the long-time behavior 
of the energy density \eqref{eq:en-dens}.  
To this end, differentiate both sides of 
$\Sigma\circ F^j = R_{j\sigma} \circ \Sigma$ 
(which is \eqref{eq:SigmaR} iterated $j$ times) 
to obtain 
\[
\Sigma'(F^j(t)) = \frac{\Sigma'(t)}{(F^j)'(t)} \ .
\]
On the other hand, taking the Schwarzian derivative 
of the same relationship and using \eqref{eq:S-GoH} 
with $G=\Sigma$, $H=F^j$, we have 
\[
\scrS_\Sigma(F^j(t)) = \frac{1}{\left[(F^j)'(t)\right]^2} 
        \left[ \scrS_\Sigma(t) - \scrS_{F^j}(t) \right] \ .
\]
These expressions yield 
\begin{equation}  \label{eq:Phi-Fj}
\Phi(F^j(t)) = \frac{1}{\left[(F^j)'(t)\right]^2} 
        \left[ \Phi(t) - \scrS_{F^j} (t) \right] 
\ .
\end{equation}
Using this equation, we can compute the energy 
of the field at an arbitrary space-time point 
if the function $\Sigma(t)$ is known for $t\in[\bar t, F(\bar t))$ 
where $\bar t$ is an arbitrary value; 
in particular, if for $t<0$ the mirrors are at rest, 
we can take $\Sigma(t)$ as in the discussion 
after \eqref{eq:a-smooth} 
(there we used the smoothing factor 
$1 - \e^{-t^4}$ because $\scrS_\Sigma$ 
contains third derivatives of $\Sigma$).  
In the case of classical EM field, 
the evolution of the energy density is similar to \eqref{eq:Phi-Fj} 
except for the term $\scrS_{F^j} (t)$, 
which corresponds to the purely quantum effect 
of creation of field by the motion of the mirror.  
Of course, in absence of electromagnetic field 
in the cavity at $t=0$, the classical energy is zero 
for all $t\geq 0$, while the energy of the quantum field 
is non-zero for $t>0$ even if at $t=0$ it was zero.

The composition rule \eqref{eq:S-GoH} implies 
\[
\scrS_{F^j}(t) = \sum_{k=0}^{j-1} \scrS_F(F^k(t)) \,\, [(F^k)'(t)]^2 
= \sum_{k=0}^{j-1} \scrS_F(F^k(t)) \prod_{i=0}^{k-1} [F'(F^i(t))]^2 \ ,
\]
which allows us to rewrite equation \eqref{eq:Phi-Fj} as 
\begin{equation}  \label{eq:Phi-Fj-iter}
\Phi(F^j(t)) = \frac{\Phi(t)}{\left[(F^j)'(t)\right]^2} 
        - \sum_{k=0}^{j-1} 
          \frac{\scrS_F(F^k(t))}{F'(F^{j-1}(t))^2\,\cdots\,F'(F^k(t))^2}
\ .
\end{equation}


\subsection{Physical mechanism of the energy changes}  
        \label{subsec:en-change}

In this subsection, we give 
a transparent physical interpretation 
of the terms in the right-hand side 
of \eqref{eq:Phi-Fj-iter}.  
Namely, the first term in the right-hand side of
\eqref{eq:Phi-Fj-iter} is the initial energy density 
amplified in the $j$ reflections from the moving mirror 
between $t$ and $F^j(t)$, while the term with summation index $k$ 
in the sum corresponds to the field created due to the motion 
of the mirror at time $F^k(t)$ and subsequently amplified 
at each of the following reflections.  
To prove this, 
we will use the following result 
concerning the energy density emitted 
by a single moving mirror.  
Fulling and Davies \cite{FullingDavies1976} 
proved that if a perfect mirror is moving in vacuum 
according to $x=a(t)$, then the regularized energy density 
to the right of the mirror (i.e., for $x>a(t)$) 
is given by 
\[
\langle T_{00}(t,x) \rangle_\mathrm{reg} 
= -\frac{1}{24\pi} \scrS_F(t-x) 
= -\frac{1}{24\pi} \scrS_F\bigl((\mathrm{Id}-a) (t_e)\bigr) \ ,
\]
where $F$ is given by \eqref{eq:F}, 
and $t_e$ is the time of emission.  
To the left of the mirror (for $x<a(t)$), 
the energy density is given by 
\begin{equation}  \label{eq:en-left}
\langle T_{00}(t,x) \rangle_\mathrm{reg} 
= -\frac{1}{24\pi} \scrS_{\widetilde F}(t+x) 
= -\frac{1}{24\pi} \scrS_{\widetilde F}\bigl((\mathrm{Id}+a) (t_e)\bigr) \ , 
\end{equation}
where $\widetilde F = (\mathrm{Id}-a)\circ(\mathrm{Id}+a)^{-1}$ 
is defined similarly to $F$ \eqref{eq:F}, 
but with $a$ replaced by~$-a$.  
Similarly to \eqref{eq:Theta}, 
we define the function 
$\widetilde \Theta = (\mathrm{Id}+a)^{-1}$.  

To rewrite \eqref{eq:en-left} in another form, 
we will need the following formulae:
\[
\scrS_{\mathrm{Id}\pm a}(\xi) 
        = \frac{a'''(\xi)\, [\pm 1 + a'(\xi)] - \frac32 a''(\xi)^2}
                {[1\pm a'(\xi)]^2} \ ,
\]
(which follows directly from \eqref{eq:Schw}), and 
\[
\scrS_G(G^{-1}(\xi)) = -[G'(G^{-1}(\xi))]^2 \, \scrS_{G^{-1}}(\xi) 
\]
(a consequence of \eqref{eq:S-GoH}), 
which in turn implies
\[
\scrS_{\widetilde\Theta}\bigl((\mathrm{Id}+a)(t_e)\bigr) 
= - \widetilde\Theta'\bigl((\mathrm{Id}+a)(t_e)\bigr)^2 \, 
        \scrS_{\mathrm{Id}+a}(t_e) 
= - \frac{\scrS_{\mathrm{Id}+a}(t_e)}
                {[1+a'(t_e)]^2} \ .
\]
Now we have 
\begin{eqnarray}  \label{eq:left1}
\scrS_{\widetilde F}\bigl((\mathrm{Id}+a)(t_e)\bigr) 
&=& 
\scrS_{(\mathrm{Id}-a)\circ\widetilde\Theta}\bigl((\mathrm{Id}+a)(t_e)\bigr) 
        \nonumber \\[2mm]
&=& 
\bigl[\widetilde\Theta'\bigl((\mathrm{Id}+a)(t_e)\bigr)\bigr]^2 \, 
        \scrS_{\mathrm{Id}-a}(t_e) 
                + \scrS_{\widetilde\Theta}\bigl((\mathrm{Id}+a)(t_e)\bigr) 
        \nonumber \\[2mm]
&=& 
\bigl[\widetilde\Theta'\bigl((\mathrm{Id}+a)(t_e)\bigr)\bigr]^2 \, 
        \left[ \scrS_{\mathrm{Id}-a}(t_e) 
                - \scrS_{\mathrm{Id}+a}(t_e)\right] 
        \nonumber \\[2mm]
&=& 
-2\, \frac{a'''(t_e)\left[1-a'(t_e)^2\right] + 3a'(t_e)a''(t_e)^2}
        {\left[1-a'(t_e)\right]^4\, 
        \left[1+a'(t_e)\right]^2}\ .
\end{eqnarray}

Now we will use \eqref{eq:left1} 
to understand the physical meaning of \eqref{eq:Phi-Fj-iter}.  
For simplicity, 
we consider the evolution of the energy density 
along a characteristic which at time $t_0$ passes through 
the point $x_0$ while moving to the left.  
At time $t_+:=t_0+x_0$ this characteristic 
is reflected from the stationary mirror, 
then at time $\Theta(t_+)$ it is reflected from the moving mirror; 
the next three reflections occur at times 
$F(t_+)$, $\Theta(F(t_+))$, and $F^2(t_+)$, respectively.  
Let $t_1$ be time after $F^2(t_+)$ but before the next reflection, 
and $x_1=t_1-F^2(t_+)$ be the spatial coordinate of the characteristic 
at time $t_1$.  For the energy density 
we obtain from \eqref{eq:Phi-Fj-iter}
\begin{eqnarray}  \label{eq:example}
\langle T_{00}(t_1,x_1)\rangle_\mathrm{reg} 
&=& 
D(\Theta(F(t_+)))^2 \, D(\Theta(t_+))^2 \, 
        \langle T_{00}(t_0,x_0)\rangle_\mathrm{reg} 
        \nonumber \\[2mm]
& &     + D(\Theta(F(t_+)))^2 \, \frac{1}{24\pi}
                \frac{\scrS_F(t_+)}{F'(t_+)^2} 
        + \frac{1}{24\pi}\frac{\scrS_F(F(t_+))}{F'(F(t_+))^2} 
\ .
\end{eqnarray}

The first term in the right-hand side of \eqref{eq:example} 
is the initial energy density 
$\langle T_{00}(t_0,x_0)\rangle_\mathrm{reg}$ 
amplified by the factor of $D(\Theta(F(t_+)))^2=[F'(t_+)]^{-2}$ 
at the reflection from the moving mirror at time $\Theta(t_+)$, 
and by the factor of 
$D(\Theta(F(t_+)))^2=[F'(F(t_+))]^{-2}$ 
at the next reflection from the moving mirror at time
$\Theta(F(t_+))$.  

The second term in the right-hand side of \eqref{eq:example} 
is the energy density 
created at the reflection of the characteristic 
from the moving mirror at time $\Theta(t_+)$, 
and consequently amplified by the factor of $D(\Theta(F(t_+)))^2$ 
at the next reflection from the moving mirror.  
Indeed, we have 
\begin{eqnarray*}
\scrS_F(t_+) 
&=& 
\scrS_{(\mathrm{Id}+a)\circ\Theta}(t_+) 
= [\Theta'(t_+)]^2 \, \scrS_{\mathrm{Id}+a}(\Theta(t_+)) 
        + \scrS_\Theta(t_+) \\[2mm]
&=& [\Theta'(t_+)]^2 \, 
        \left[ \scrS_{\mathrm{Id}+a}(\Theta(t_+)) 
                - \scrS_{\mathrm{Id}-a}(\Theta(t_+))\right] \\[2mm]
&=& 
2\, \frac{a'''(\xi)\left[1-a'(\xi)^2\right] + 3a'(\xi)a''(\xi)^2}
        {\left[1-a'(\xi)\right]^4\, 
        \left[1+a'(\xi)\right]^2}\Biggr|_{\xi=\Theta(t_+)} \ .  
\end{eqnarray*}
Together with \eqref{eq:left1}, 
this equality implies 
\begin{eqnarray}  \label{eq:LR}
\frac{1}{24\pi} \frac{\scrS_F(t_+)}{F'(t_+)^2} 
&=& 
\frac{1}{12\pi}\,\frac{a'''(\xi)\left[1-a'(\xi)^2\right] 
        + 3a'(\xi)a''(\xi)^2}
        {\left[1-a'(\xi)\right]^2\, 
        \left[1+a'(\xi)\right]^4}\Biggr|_{\xi=\Theta(t_+)} 
        \nonumber \\[2mm]
&=& 
- \frac{1}{24\pi} 
\scrS_{\tilde F}\bigl((\mathrm{Id}+a)(\Theta(t_+))\bigr) \ , 
\end{eqnarray}
and a comparison with \eqref{eq:LR} 
proves the correctness of our interpretation 
of the term $\frac{1}{24\pi} \frac{\scrS_F(t_+)}{F'(t_+)^2}$.  

Similarly, the third term in the right-hand side 
of \eqref{eq:example} is the energy density 
created by the moving mirror at time $\Theta(F(t_+))$.  

This discussion elucidates the difference 
between the classical and the quantum cases 
-- in the classical case 
the moving mirror amplifies 
the wave packets by squeezing them, 
while in the quantum case the moving mirror 
not only amplifies the already existing wave packets, 
but also creates new field which subsequently 
is amplified at each reflection from the moving mirror.


\subsection{Resonant amplification in a periodically pulsating cavity}  
        \label{subsec:ampl-period}

Now we will consider the particular case 
of periodic motion of the mirror 
when the corresponding CM $f$ 
has rational rotation number, $\tau(f) = \sigma = p/q$, 
i.e., is in $p/q$ phase locking.  
In this case the classical EM field in the cavity 
develops wave packets whose number can be anywhere 
between 1 and $q$ depending on the initial conditions.  
The wave packets become narrower at each reflection, 
and their energy increases at each reflection.  

In the quantum case, the motion of the mirror itself 
creates energy which is subsequently concentrated 
in narrow wave packets.  
If the acceleration of the the mirror is not zero 
except at isolated times, 
then the mirror is emitting energy all the time, 
and the number of wave packets developed 
is {\em exactly}~$q$.  

In the case of $p/q$ phase locking, 
the CM $f$ generically has an attracting $q$-periodic orbit, 
which corresponds to the times of reflection of the 
attracting characteristics.  
If $t^*$ is such a time 
(i.e., if the fractional part of $t^*$ 
belongs to the attracting $q$-periodic orbit of $f$), then 
\[
F^q(t^*)=t^*+p \ , 
\quad 
0<(F^q)'(t^*) <1 \ , 
\]
which implies that asymptotically 
the cumulative Doppler factor 
$\calD_q=[(F^q)'(t^*)]^{-1}$ 
(see \eqref{eq:D-q}) is greater than~$1$.  
In the case of $p/q$ phase locking, 
\eqref{eq:Phi-Fj-iter} implies 
(for any integer $n$) 
\begin{equation}  \label{eq:Phi-Fnq}
\Phi(F^{nq}(t)) 
= 
\frac{\Phi(t)}{\left[(F^{nq})'(t)\right]^2} 
- \sum_{j=0}^{n-1} 
        \frac{\scrS_{F^q}(F^{jq}(t))}
                {\left[(F^{(n-j)q})'(F^{jq}(t))\right]^2} 
\ .
\end{equation}
If $t^*$ is a time of reflection 
of an attracting characteristic, 
\eqref{eq:Phi-Fnq} yields 
\begin{eqnarray}  \label{eq:reson-en}
\Phi(t^*+np) 
&=& 
        \calD_q^{2n} \left( \Phi(t^*) 
        - \scrS_{F^q}(t^*) \sum_{j=0}^{n-1} \calD_q^{-2j} \right) 
        \nonumber \\ 
&=& 
        \calD_q^{2n} \left( \Phi(t^*) 
        - \frac{1-\calD_q^{-2n}}{1-\calD_q^{-2}} 
                \scrS_{F^q}(t^*) \right)  \ .
\end{eqnarray} 
The applicability of this expression 
is not restricted to times like $t^*$ 
-- since asymptotically all characteristics are very close 
to the attracting ones, 
\eqref{eq:reson-en} gives approximately 
the asymptotic behavior of the energy density 
of the wave packets.  
Parenthetically, we remark that 
using this expression, one can prove that 
for motion of the mirror with parameters 
corresponding to the ends of the phase-locking intervals 
(i.e., the ends of the horizontal parts 
of the graph in Figure~\ref{fig:devil}(b)), 
the energy density grows not exponentially, 
but polynomially:
\[
\lim_{\calD_q\to1^+} \Phi(t^*+np) 
= \Phi(t^*) - n \scrS_{F^q}(t^*) \ .
\]

The energy of the $j$th wave packet at time~$t$ 
is given by 
\[
\calE_j(t) 
        = \int_{\Delta_j(t)}\langle T_{00}(t,x)\rangle_{\mathrm{reg}}
                \, \mathrm{d} x 
        = -\frac{1}{24\pi} \int_{\Delta_j(t)} 
                \left[ \Phi(t-x) + \Phi(t+x) \right]
                \, \mathrm{d} x  \ ,
\]
where $\Delta_j(t)$ is the support of the $j$th wave packet 
at time~$t$.  
Recalling that at reflection at time $t$ the width of the wave packet 
decreases $D(t)$ times, we obtain that 
asymptotically the energy of the field in the cavity 
increases exponentially: for large $t$, 
\[
\calE_\mathrm{total}(t) \sim \mathrm{const} \cdot \calD_q^{t/p} \ .
\]

\begin{figure}
  \centerline{\psfig{file=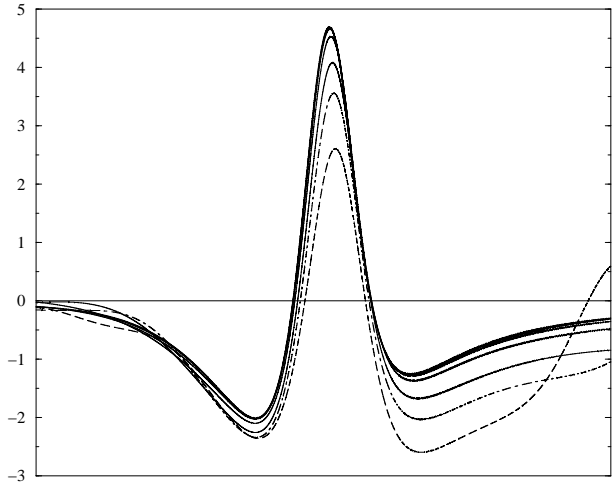,%
                     width=0.39\textwidth,angle=-90}
        }
 \caption{\label{fig:shape-packet}
        Rescaled ``shape'' of a wave packet at times 
        2 (dashed line), 3 (dot-dashed line), 
        4, 6, 8, 10, 12 (solid lines) 
        (see the text).  
   }
\end{figure}

The energy density of the wave packets changes after 
each reflection from the moving mirror, 
but after appropriate rescaling, 
the its ``shape'' at times $np$ as $n\to\infty$ 
tends to a some constant profile 
which depends on the motion of the mirror.  
In Figure~\ref{fig:shape-packet}, 
we show the evolution of the ``shape'' of a wave packet 
for mirror's motion same as in Figure~\ref{fig:graph-S}(a).  
In the figure we show the rescaled energy density, 
$\calD_q^{-2n}\cdot\langle T_{00}(n,x)\rangle_{\mathrm{reg}}$, 
on the vertical axis versus 
the shifted and rescaled spatial coordinate 
$\calD_q^n\cdot(x-x_n^*)$ on the horizontal axis, 
at times $n$ for several values of~$n$; 
here $x_n^*$ is the spatial coordinate of the attracting 
characteristic corresponding to this packet, 
at time~$n$.


\section{Concluding remarks}

The power of the methods of theory of dynamical systems 
is due to their generality.  
The predictions we have made about the behavior 
of the field in the cavity are applicable to any 
motion of the mirror, not only to particular examples.  
Within our approach, 
we gave a complete classification of the 
possible resonances (phase locking) in the system, 
predicted that, generically, small detuning does 
not destroy the resonance, 
gave a simple explanation of the squeezing of the wave packets, 
interpreted the origin of the different contributions 
to the energy density in the cavity.  
We would also like to emphasize 
that our technique is non-perturbative.  

In the case of resonance, the standard numerical methods for 
solving partial differential equations would be very difficult to 
apply because of the concentration of the field in narrow packets.  
The proposed method, however, 
relies on iterating one-dimensional maps, 
so that resonances do not present any additional difficulty.  
The computer programs used to produce the pictures in this paper 
took minutes to run on a PC.  

Our methodology easily generalizes (see our paper \cite{PLV03}) 
to the case of quasiperiodic motion 
of the mirror, and the case of two moving mirrors 
(studied previously by Ji \etal\ \cite{JJS98}, 
Dodonov \cite{Dod98}, 
Dalvit and Mazzitelli \cite{DM99-PRA}, 
Li and Li \cite{LiL02}).  
Recently, similar ideas from dynamical systems 
have been applied to the study 
of waves in a fluid in two-dimensional basin 
by Manders \etal\ \cite{MandersDM03}.

Interestingly, the behavior of the field of the cavity (described by a 
partial differential equation) is easier to analyze than the behavior 
of a particle bouncing back and forth between two perfectly 
reflecting walls (assuming that the reflections are perfectly elastic).  
The latter system, suggested by Fermi \cite{Fermi49} as a possible 
mechanism for acceleration of the particles in the cosmic rays, 
reveals a much richer dynamical behavior (see, e.g., 
the book of Lichtenberg and Lieberman \cite{LichLieb92}).

There are many questions that deserve a further study.  
An interesting question is whether 
theory of dynamical systems can be applied 
to the case of a constant-length cavity 
filled with dielectric with changing properties.  
Another problem is the absence of 
resonances of certain type (noticed in our paper \cite{LP99} 
and discussed by W\c{e}grzyn \cite{Wegrzyn04-PLA}) -- is it generic, 
and which resonances are forbidden?  
Can methods of dynamical systems be used 
to study the problem in higher dimensions?  
Can our methodology be applied 
to the recent suggestion by Jaffe and Scardicchio \cite{JaffeS04} 
to apply methods of geometric optics to the study 
of Casimir effect?  
Can similar methods be applied to other fields 
(see, e.g., the study of a classical massive field 
in a pulsating resonator by 
Dittrich and Duclos \cite{DittrichD02})?  

The appearance of the Schwarzian derivative hints 
at possible deeper connections between the quantum problem of a 
moving mirror, partial differential equations, 
and dynamical systems.


\section*{Acknowledgments}

I would like to express my gratitude to Rafael de la Llave 
who directed my attention to the classical aspects 
of the problem considered in this paper, 
which resulted in our paper \cite{LP99} 
and motivated other collaborative research \cite{LP02}.  
I had a very pleasant collaboration 
with him and John Vano on the generalization 
of our method to the case of quasiperiodic motion 
of the mirror \cite{PLV03}.  
I would like to acknowledge fruitful conversations 
with Gautam Bharali and Michael Bolt.  
My research was partially supported by the Rackham Faculty 
Fellowship of the Rackham Graduate School, University of Michigan.  

Last but not least, 
I would like to thank the organizers of this topical issue, 
Gabriel Barton, Victor V.\ Dodonov, and Vladimir I.\ Man'ko, 
for the invitation to submit a paper.


\section*{References}
\bibliographystyle{plain} 
\bibliography{quantum_cavity}

\end{document}